\documentclass[twocolumn]{aastex63}

\usepackage{amsmath}
\usepackage{pdflscape}
\usepackage{pdflscape}
\usepackage{afterpage}

\makeatletter
\AtBeginDocument{\let\LS@rot\@undefined}
\makeatother

\accepted{August 27, 2021}
\submitjournal{ApJ}

\shorttitle{``Worst-Case" Quasar Micro-Lensing}
\shortauthors{Weisenbach, Schechter, and Pontula}

\begin{document}

\title{``Worst-Case" Micro-Lensing in the Identification and Modeling of Lensed Quasars}

\correspondingauthor{Luke Weisenbach}
\email{weisluke@alum.mit.edu}

\author[0000-0003-1175-8004]{Luke Weisenbach}
\affiliation{MIT Department of Physics \\
Cambridge, MA 02139 USA}

\author[0000-0002-5665-4172]{Paul L. Schechter}
\affiliation{MIT Department of Physics \\
Cambridge, MA 02139 USA}
\affiliation{MIT Kavli Institute for Astrophysics and Space Research \\
Cambridge, MA 02139 USA}

\author{Sahil Pontula}
\affiliation{MIT Department of Physics \\
Cambridge, MA 02139 USA}

%% Mark off the abstract in the ``abstract'' environment. 
\begin{abstract}

Although micro-lensing of macro-lensed quasars and supernovae provides unique opportunities for several kinds of investigations, it can add unwanted and sometimes substantial noise.  While micro-lensing flux anomalies may be safely ignored for some observations, they severely limit others. ``Worst-case" estimates can inform the decision whether or not to undertake an extensive examination of micro-lensing scenarios. Here, we report ``worst-case" micro-lensing uncertainties for point sources lensed by singular isothermal potentials, parameterized by a convergence equal to the shear and by the stellar fraction. The results can be straightforwardly applied to non-isothermal potentials utilizing the mass sheet degeneracy. We use micro-lensing maps to compute fluctuations in image micro-magnifications and estimate the stellar fraction at which the fluctuations are greatest for a given convergence. We find that the worst-case fluctuations happen at a stellar fraction $\kappa_\star=\frac{1}{|\mu_{macro}|}$. For macro-minima, fluctuations in both magnification and demagnification appear to be bounded ($1.5>\Delta m>-1.3$, where $\Delta m$ is magnitude relative to the average macro-magnification). Magnifications for macro-saddles are bounded as well ($\Delta m > -1.7$). In contrast, demagnifications for macro-saddles appear to have unbounded fluctuations as $1/\mu_{macro}\rightarrow0$ and $\kappa_\star\rightarrow0$. 

\end{abstract}

%% Keywords should appear after the \end{abstract} command. 
%% See the online documentation for the full list of available subject
%% keywords and the rules for their use.

\section{Introduction} \label{sec:intro}

Gravitational lensing has emerged as a powerful tool to probe the potentials of distant galaxies and galaxy clusters, and has further been used to help constrain models of dark matter and dark energy. Lensed quasars have been particularly prominent, and the number of known quadruply lensed quasars has increased by roughly 50\% since the first data release from the Gaia satellite \citep{2016A&A...595A...2G}. Approximately 20 quadruply lensed quasars have been discovered using Gaia DR1 and DR2, often in concert with other catalogs. 

The Gaia GraL group \citep{2018A&A...616L..11K} identified 80,000 quartets of candidate point sources for consideration as possible quadruply lensed quasars. The positions and fluxes were analyzed, with great success, using ``Extremely Randomized Trees" \citep{2019A&A...622A.165D}, with twelve out of thirteen known quadruply lensed quasars with four catalogued images ranked in the top 0.6\%. The authors attribute the ranking of the thirteenth, 3 times further down the list, to micro-lensing of the quasar images by stars in the lensing galaxy, a phenomenon that had not been taken into account in their training set of roughly $10^8$ systems. 

Micro-lensing helps place constraints on the smooth (dark) matter content of the lensing galaxy. Previous work has shown that, for highly magnified macro-images, the fluctuations in flux ratio due to micro-lensing often increases with the introduction of dark matter to the lens model \citep{2002ApJ...580..685S}. Furthermore, it is expected that these micro-lensing fluctuations are larger for saddle-points as compared to minima \citep{1995ApJ...443...18W}, which large-scale parameter studies of micro-lensing magnification probability distributions have confirmed \citep{1992ApJ...386...19W, 1995MNRAS.276..103L, 2013MNRAS.434..832V}. Contingent upon available resources, it would be possible to include the effects of micro-lensing in the training set of the Extremely Randomized Trees, particularly with the public availability of the GERLUMPH suite of simulations \citep{2014ApJS..211...16V}. Here we describe an approximate alternative that could be readily adapted to lens searches, which inevitably involve expectations for lensed flux ratios. 

Briefly, we assume a convergence and shear appropriate to a singular isothermal elliptical potential -- a model that works reasonably well for most know quadruply lensed quasar systems -- and search for the stellar contribution to the convergence that produces the largest micro-lensing fluctuations for a point source. We call this ``worst-case" micro-lensing, and give the 95\% confidence range for the fluctuations. One can then use these ranges (accounting in some way for the likelihood of the worst case stellar fraction) to assign a likelihood to a discrepant source flux. The modeling of Gaia quartets is just one of many lens modeling problems where worst case estimates could prove useful. One needn't completely discount the observed fluxes from quasar images if expected fluctuations are small, as for low magnification images (particularly macro-minima, which as stated previously suffer less from such fluctuations than macro-saddles). 

The fluctuations discussed within this work are for that fraction of the source that can be treated as pointlike. Our discussions are not relevant to the flux from more extended regions of emissions corresponding to longer wavelength observations, which have narrower magnification distributions \citep{2007MNRAS.381.1591B}. 

The organization of this paper is as follows: In \S\ref{sec:background}, we give a brief introduction to gravitational lensing, macro-lensing models, and the phenomenon of micro-lensing. In \S\ref{sec:method}, we describe the important components of our methodology, including the micro-lensing maps upon which our analyses depended. \S\ref{sec:results} shows our results, including plots of worst-case fluctuations. A set of systems which showcase ``worst-case'' conditions are discussed in \S\ref{sec:observed_worst_cases}. Conclusions are presented in \S\ref{sec:conclusions}.

\section{Background} \label{sec:background}

\subsection{Gravitational Lensing Phenomenon}

Gravitational lensing is a direct consequence of general relativity. Light passing near a massive object is deflected, much as a light ray obeying Fermat’s principle of least time refracts on passing through an optical lens. Gravitational lensing can be succinctly described using the time delay surface \begin{equation}
    t = \frac{1+z_d}{c}\Big[\frac{D_d D_s}{2D_{ds}}\Big(\frac{\boldsymbol{\xi}}{D_d} - \frac{\boldsymbol\eta}{D_s}\Big)^2 - \Psi(\boldsymbol\xi)\Big]
\end{equation} where $z_d$ is the redshift of the lens, $D_d$, $D_s$, and $D_{ds}$ are the angular diameter distances from the observer to lens, observer to source, and lens to source respectively, $\boldsymbol\eta$ is the position in the source plane, $\boldsymbol\xi$ is the position in the image plane, and \begin{equation}
    \Psi(\boldsymbol\xi) = \frac{2}{c^2}\int_{source}^{observer}\phi(\boldsymbol\xi,l) \,\mathrm{d}l
\end{equation} is the Newtonian gravitational potential of the lens $\phi(\boldsymbol\xi,l)$ scaled and integrated along the line of sight \citep{1992grle.book.....S}. The lens equation, which can be simply written as \begin{equation}
    \nabla t = 0
\end{equation} or more fully as \begin{equation}
    \boldsymbol\eta = \frac{D_s}{D_d}\boldsymbol{\xi} - D_{ds}\nabla\Psi(\boldsymbol\xi),
\end{equation} provides the locations where images of the source are seen -- at stationary points of the time delay surface, which can be either minima, maxima, or saddle-points. The magnifications of the images are inversely proportional to the curvature of the time delay surface.

The lens equation can be non-dimensionalized into \begin{equation}
    \boldsymbol y = \boldsymbol x - \nabla\psi(\boldsymbol x)
\end{equation} where $\bf y$ is the position in the source plane, $\bf x$ is the position in the image plane, and $\psi$ is the gravitational potential. The magnifications of the images are then \begin{equation}
    \mu =  1 / \det\Big(\frac{\partial\boldsymbol y}{\partial\boldsymbol x}\Big),
\end{equation} with the sign of the magnification determining the parity of the image. Saddle-points have $\mu<0$, while minima and maxima have $\mu>0$. Furthermore, minima are always magnified, whereas saddles and maxima can be demagnified. 

\begin{figure*}
    \centering
    \includegraphics[width=0.45\textwidth]{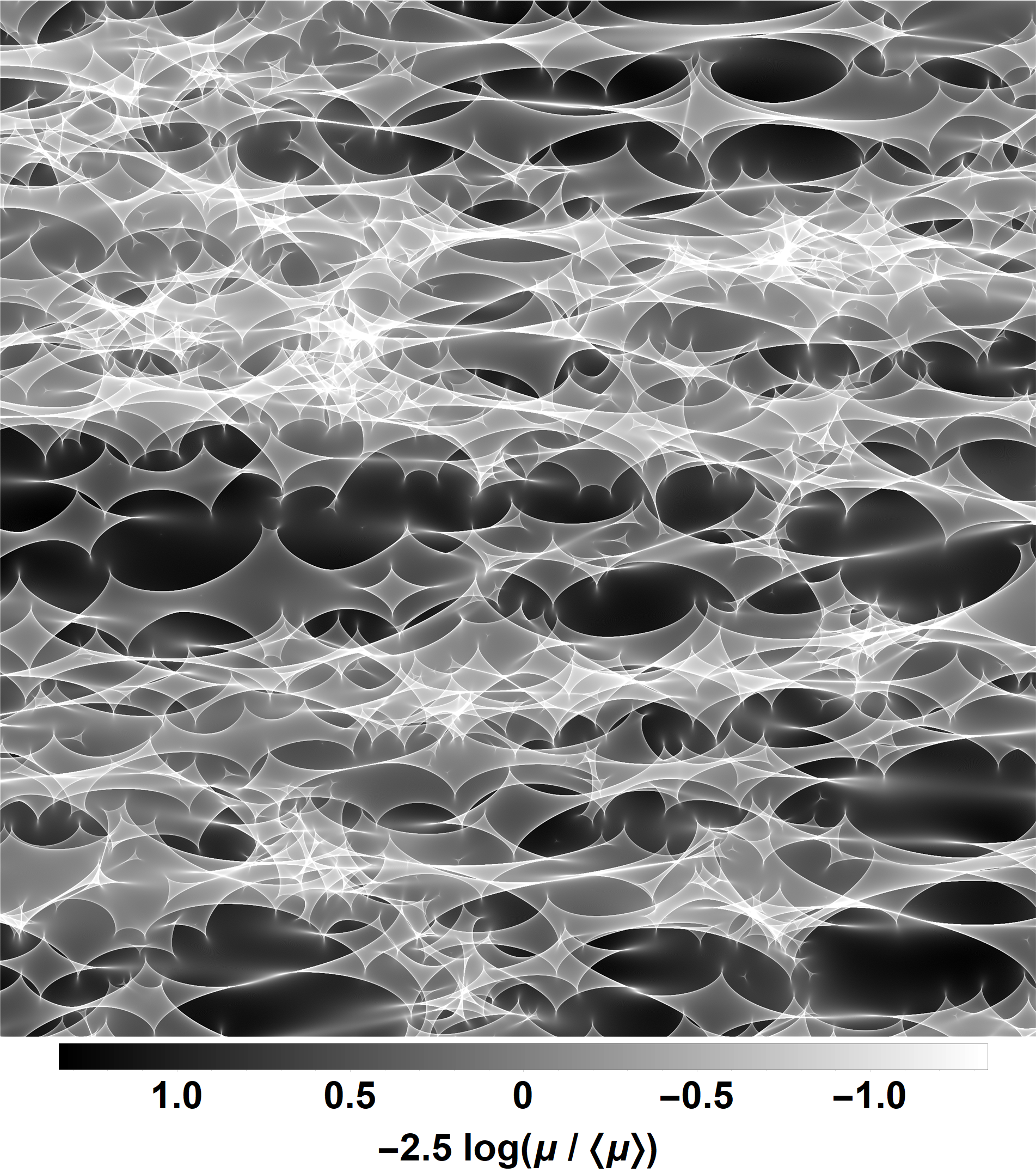}
    \includegraphics[width=0.45\textwidth]{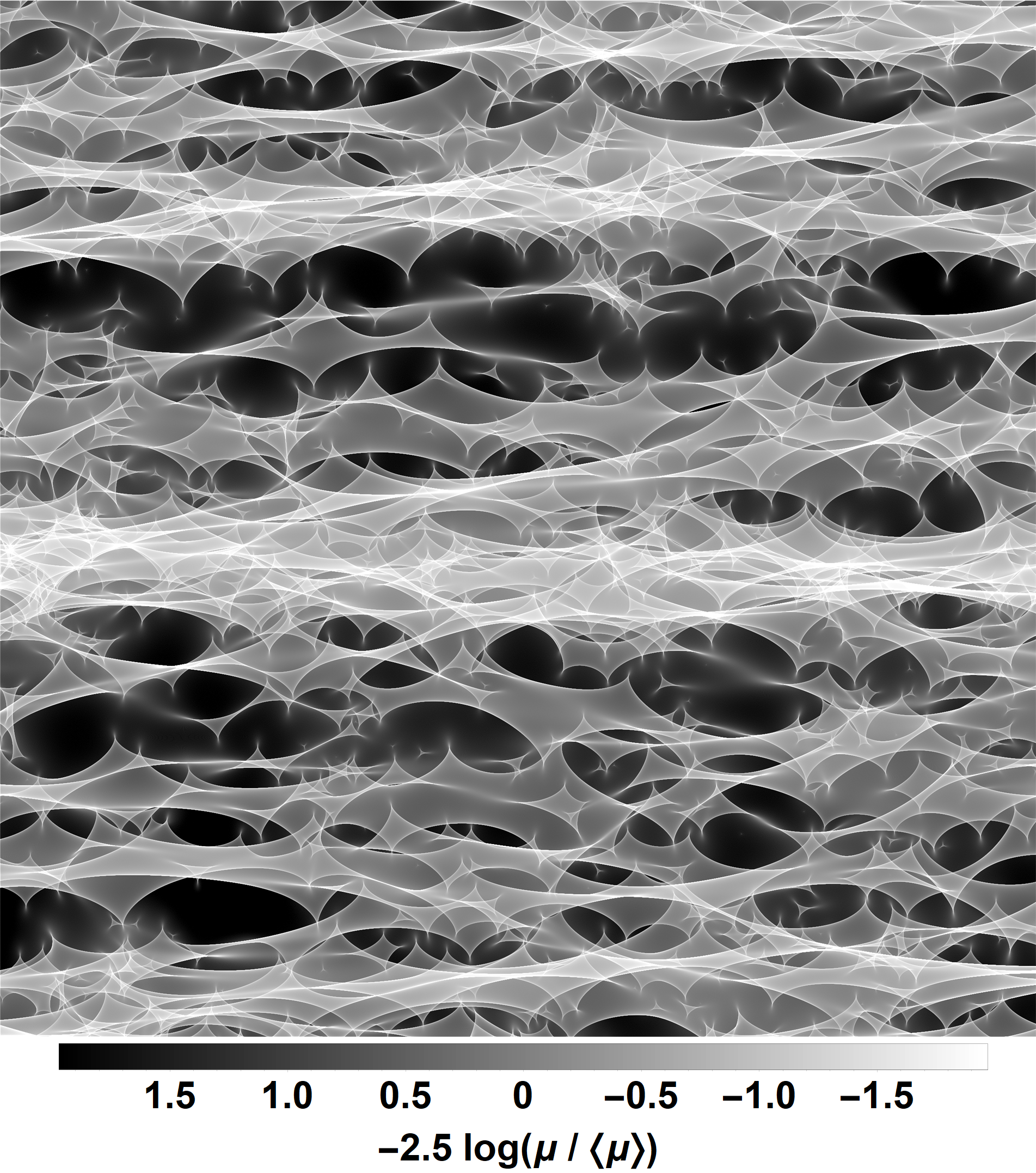}\\
    \vspace{0.3cm}
    \includegraphics[width=0.45\textwidth]{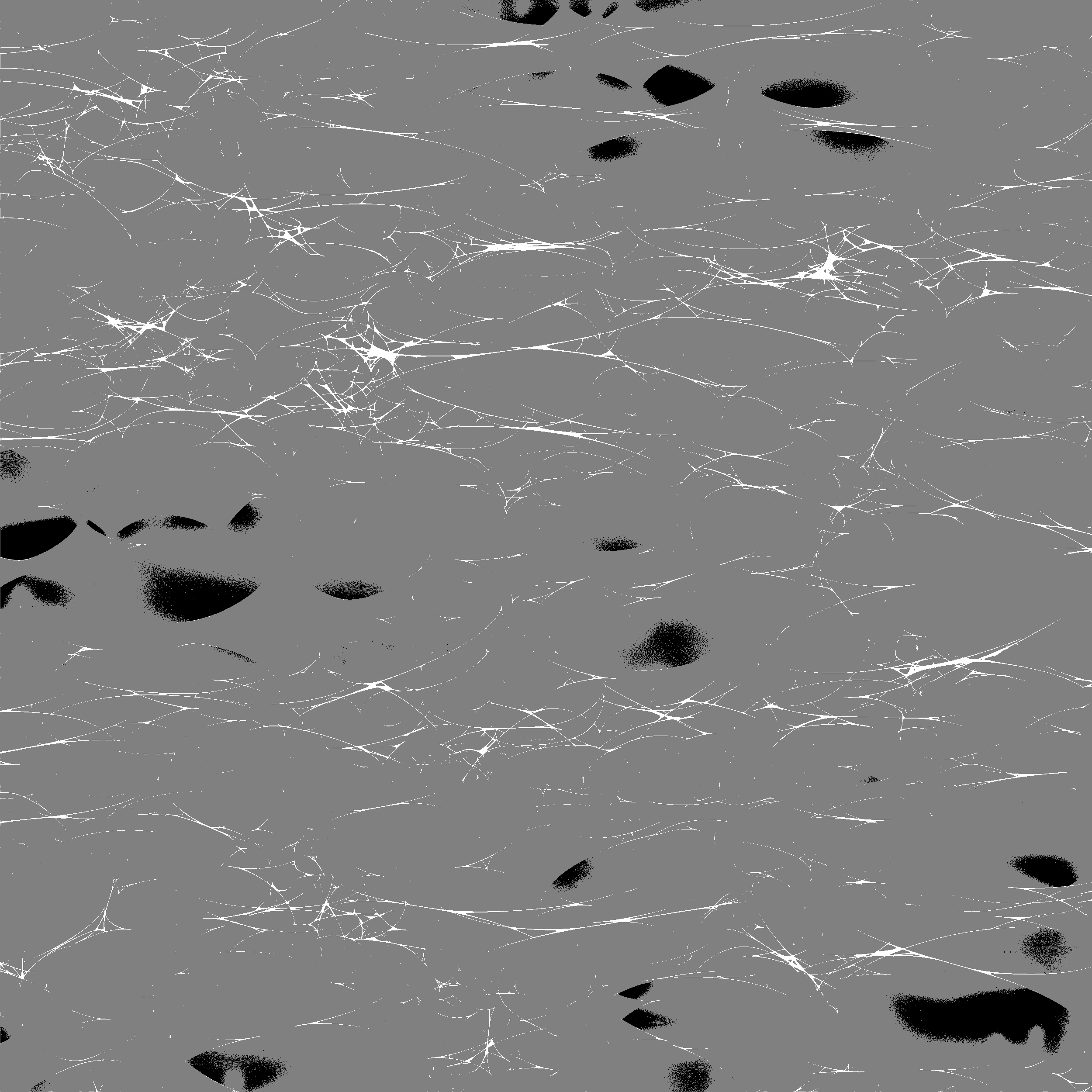}
    \includegraphics[width=0.45\textwidth]{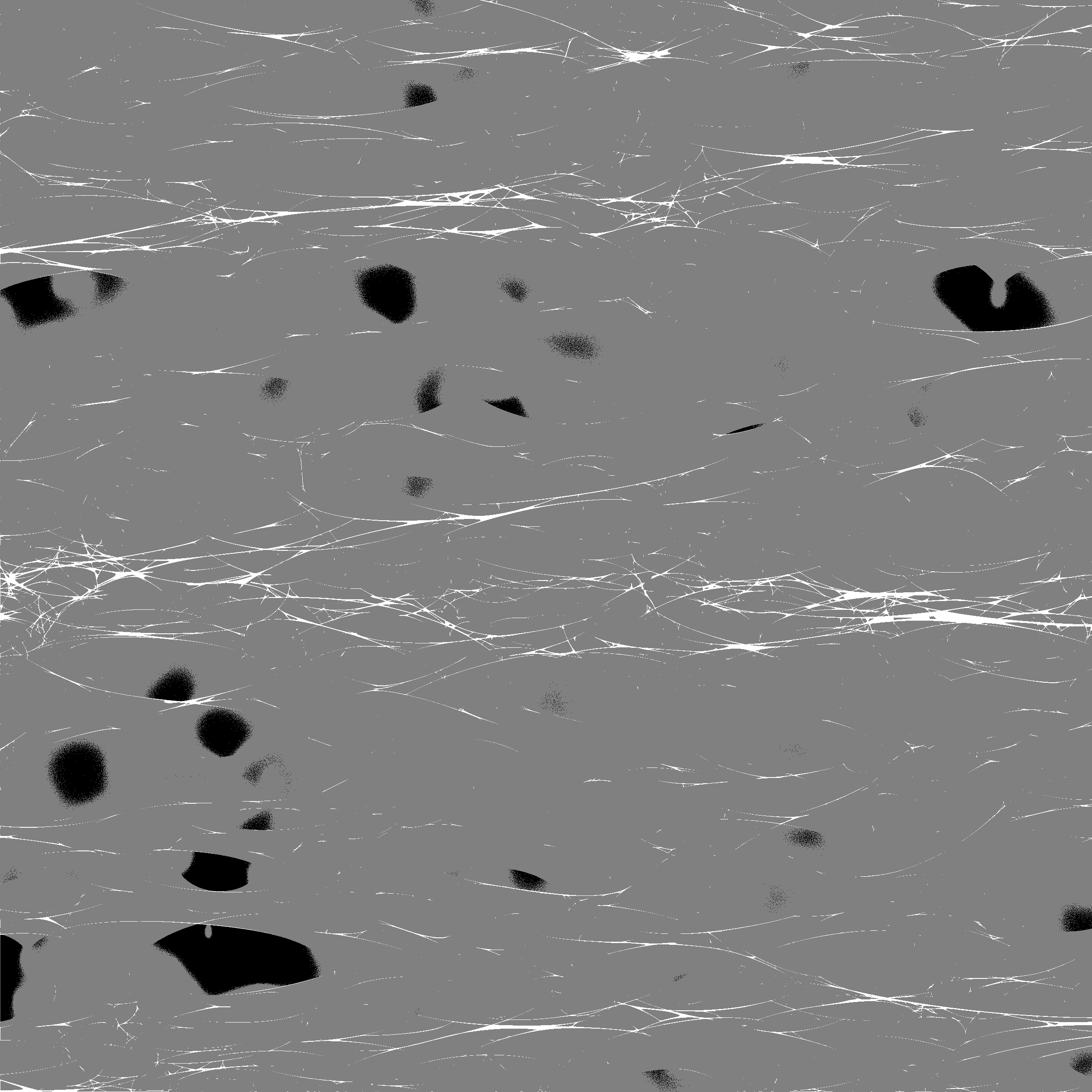}\\
    \vspace{0.2cm}
    \includegraphics[width=0.45\textwidth]{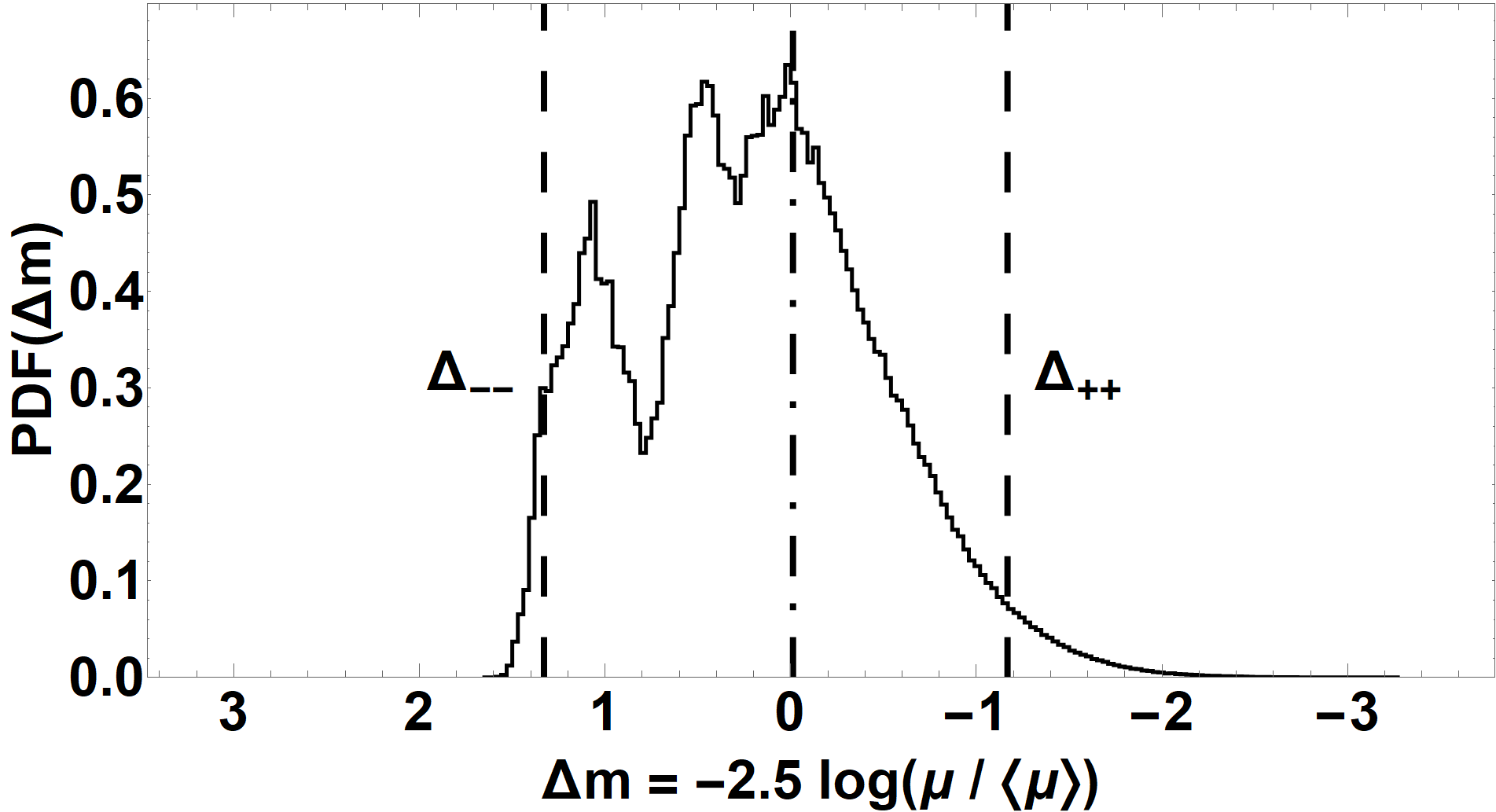}
    \includegraphics[width=0.45\textwidth]{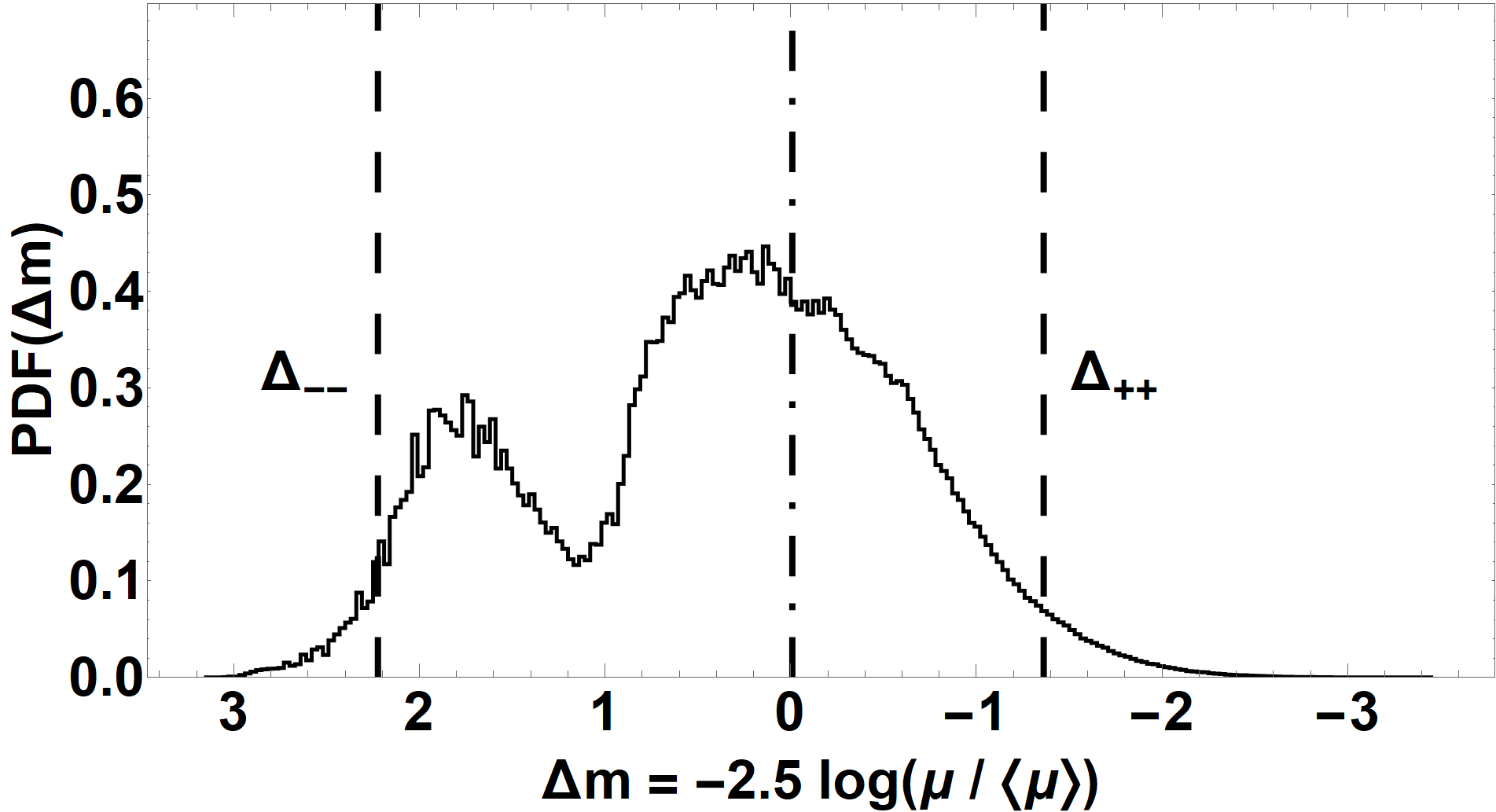}
    \caption{Top: Example magnification maps for (left) $\kappa=\gamma=0.45$, $s_\star=0.5$ and (right) $\kappa=\gamma=0.55$, $s_\star=0.5$. The physical size of the maps is $25\theta_E$ per side. Middle: The magnification maps converted to very low contrast. Black pixels denote $\mu<\Delta_{--}$, white pixels denote $\mu>\Delta_{++}$, and gray pixels cover everything between. Bottom: Histograms for the magnification maps. The left and right dashed lines in each mark the values of $\Delta_{--}$ and $\Delta_{++}$ respectively, while the central dot-dashed line marks the magnitude of the simulated average magnification. Magnitude bins have a width of $\Delta m=0.03$ mag.}
    \label{fig:mag_map}
\end{figure*}

\subsection{Quadruple Lenses and Macro-lensing Models}

Our analysis is appropriate to lensed sources with any number of images, but as quadruply lensed sources have greater redundancy, we concentrate on these. Quadruply lensed quasars play an important role in extragalactic astrophysics, helping elucidate the potentials, stellar content, and dark matter content of the lensing galaxy as well as structural properties of the background source \citep{2004IAUS..220..103S, 2008MNRAS.391.1955B}. Furthermore, quadruple lenses are of great importance in cosmology because they help constrain the values of parameters such as the Hubble constant \citep{2013ApJ...766...70S}. Lensing galaxies are often modeled as singular isothermal elliptical potentials (SIEP) \citep{1987ApJ...312...22K}, for which observable images are seen at saddle-points and minima of the light travel time surface. An image also forms at the maximum near the center of the lens’ matter distribution, but the singular nature of the SIEP model at this location means that this image is infinitely demagnified. This model has been reasonably successful at predicting the image positions \citep{2019ApJ...876....9S}. The magnifications of the macro-images can be computed as as \begin{equation}\label{eq:macro_mag}
    \mu_{macro} = \frac{1}{(1-\kappa)^2 - \gamma^2}
\end{equation} where $\kappa$ denotes the effective convergence of the lens at the position of a macro-image and $\gamma$ denotes its shear. $\kappa$ is related to the gravitational potential through the two-dimensional Poisson equation \begin{equation}
    \kappa(\boldsymbol x) = \frac{1}{2}\nabla^2\psi(\boldsymbol x) = \frac{1}{2}\big(\psi_{11}+\psi_{22}\big),
\end{equation} while $\gamma=\sqrt{\gamma_1^2+\gamma_2^2}$ with \begin{equation}
    \gamma_1 = \frac{1}{2}\big(\psi_{11}-\psi_{22}\big), \quad \gamma_2 = \psi_{12}=\psi_{21}.
\end{equation} Minima of the time delay surface occur when $1 - \kappa - \gamma > 0$, saddle-points occur when $1 - \kappa - \gamma < 0$, and maxima occur when $1 - \kappa + \gamma > 0$. 

For isothermal gravitational potentials, we further have $\kappa=\gamma$, for which equation (\ref{eq:macro_mag}) reduces to \begin{equation}\label{eq:macro_mag_simplified}
    \mu_{macro} = \frac{1}{1-2\kappa}.
\end{equation}  

\subsection{Micro-lensing}

``Micro-lensing'' is distinguished from ``macro-lensing'' in that the multiple images produced by the former cannot be resolved with today's telescopes.  Both can occur simultaneously, as when a galaxy produces multiple observable images of a quasar and the stars within the galaxy produce unresolved multiples of {\it those} images.  It is useful to distinguish between micro-lensing at low optical depth, in which a single star does the lensing, and lensing at high optical depth, when a great many stars contribute to the lensing, producing a great many micro-images \citep{1986ApJ...301..503P}. The observed flux from the macro-image is then the combined flux from the micro-images. 

As the source and the micro-lensing stars move, the micro-images change in brightness, introducing fluctuations in the observed fluxes of macro-images. As one might expect, this effect depends on the stellar content of the lens. For this reason we define the stellar fraction as \begin{equation}\label{eq:s_star}
    s_\star = \frac{\kappa_\star}{\kappa}
\end{equation} where $\kappa_\star$ is the convergence due to stars, $\kappa_s$ is the convergence due to smooth matter, and $\kappa=\kappa_\star+\kappa_s$ is the total convergence. 

\section{Method} \label{sec:method}

\subsection{Micro-lensing Maps}

Our work is based on micro-lensing magnification maps created using the inverse ray-shooting technique \citep{1986A&A...164..237S, 1986A&A...166...36K}. A multitude of light rays are traced backwards with the lens equation from image plane to source plane and collected in a pixelated map. The number of rays per pixel is then proportional to the total magnification. The dimensionless lens equation is given by \begin{equation}
    \mathbf{y} = \left(
\begin{array}{cc}1-\kappa_s+\gamma & 0\\
0 & 1-\kappa_s-\gamma
\end{array}\right)\mathbf{x} - \theta_E^2\sum_{i=1}^n\frac{m_i(\bf x - \bf x_i)}{|\bf x - \bf x_i|^2}
\end{equation} where $m_i$ is the mass of a micro-lens in units of some mass $M$ that determines the Einstein radius $\theta_E$, and $\bf x_i$ is the position of a micro-lens. For all of our simulations, we took our stars to be of unit mass and used the corresponding $\theta_E$ as our unit distance. 

\begin{turnpage}
    \begin{figure*}
        \centering
        \includegraphics[width=0.48\textheight]{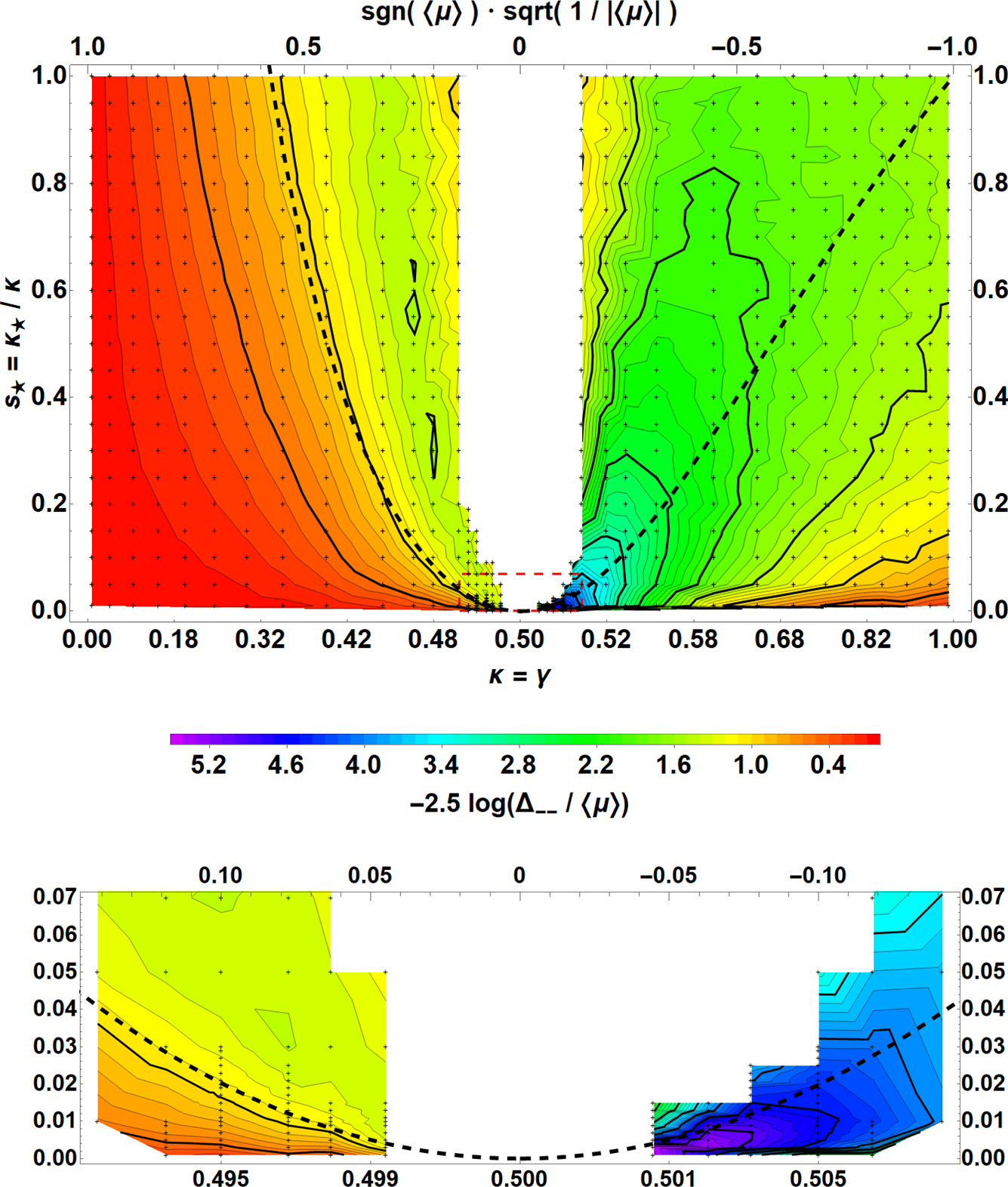}
        \hspace{0.5cm}
        \includegraphics[width=0.48\textheight]{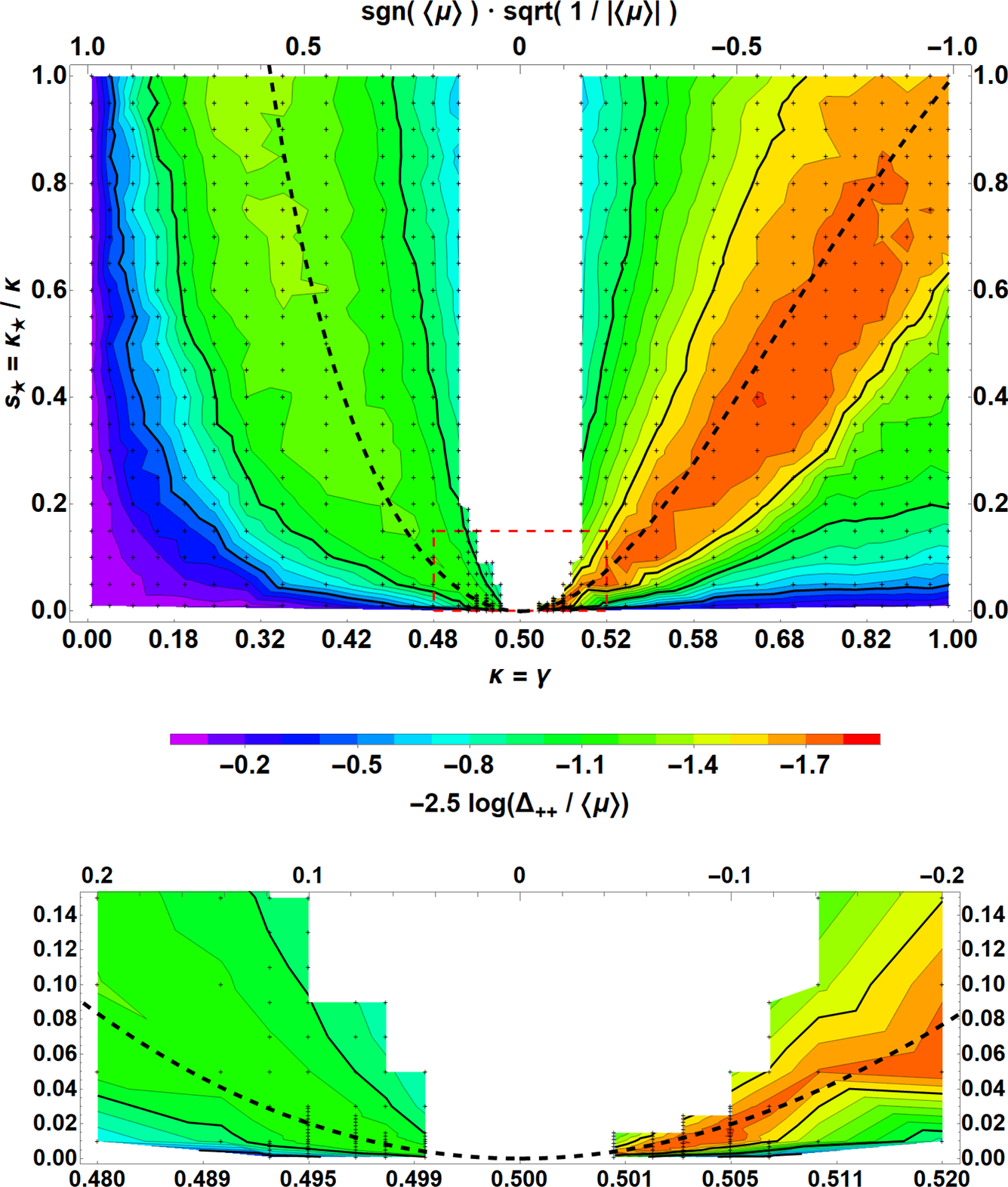}
        \hspace{0.5cm}
        \caption{Contour plots of (left) $\Delta_{--}$ and (right) $\Delta_{++}$ as functions of $\kappa=\gamma$ and $s_\star$. The ``+" symbols denote the points in the parameter space for which we ran micro-lensing simulations, while the dashed black line denotes the locus where $\kappa_\star=1/|\mu_{macro}|$. Contours are separated by 0.1 mag, with the thicker black contours occurring in 0.5 mag intervals. The top portion of each plot shows the entire parameter space covered, with the dashed red rectangle denoting the zoomed region that is shown in the bottom portion of each plot.}
    \label{fig:delta++_delta--}
    \end{figure*}
\end{turnpage}

Our investigation initially used the publicly available magnification maps of the GERLUMPH Data Release 1 \citep{2014ApJS..211...16V}, although we ultimately ended up using our own ray-shooting code that runs on a Graphics Processing Unit (GPU) to more easily examine areas of interest in the high magnification, low stellar density regime while maintaining a set of consistent simulation parameters such as map size and average ray density where possible. Our GPU implementation is neither a fully direct ray-tracing like that of \cite{2010NewA...15...16T}, nor a fully GPU parallelized tree-code like that of \cite{Teralens}. Instead, our naive implementation uses all the stars in the field to directly shoot 4 rays (one near each of the corners) in each square of a grid that constitutes the shooting region in the image plane. We calculate Taylor coefficients of the deflection angle (up to third order) within the square and use those coefficients to then shoot $\approx 700$ more rays within the square. The size of the squares depends on the number density of rays in the image plane, which itself depends on the macro-magnification $\mu_{macro}=\langle\mu\rangle$, the desired average number of rays per pixel $\langle n_{rays}\rangle$, the pixel size, and the number of rays that are shot using the Taylor coefficients. We note that the sizes of the squares in the image plane were never greater than $0.2\theta_E$ for our simulations, with the majority being less than $0.05\theta_E$ over the parameter space sampled. The top portion of Figure \ref{fig:mag_map} shows sample magnification maps created with our code. We shot rays into a square in the source plane with a side length of $25\theta_E$ and $2500$ pixels, so that each pixel is a $0.01\theta_E$ x $0.01\theta_E$ square. We additionally chose $\langle n_{rays} \rangle=1000$ for all of our simulations.

\subsection{Worst-Case Analysis}

The magnification maps show the total source magnification as a function of position. The magnification at a specific source position is the sum of the magnifications of the many micro-images, and in general will fluctuate around the average macro-magnification $\langle\mu\rangle$ as the source moves. We create histograms of the magnifications, expressed as magnitude differences from the theoretical average $\langle\mu\rangle$, using \begin{equation}
    \Delta m = -2.5\log\frac{\mu}{\langle\mu\rangle} = -2.5\log\frac{n_{rays}}{\langle n_{rays}\rangle}.
\end{equation}

We choose to examine worst-case micro-lensing by calculating the $97.5$ and $2.5$ percentiles\footnote{Chosen to reflect a quasi- 2 standard deviation spread} of the magnitude distributions; i.e., we find the ray counts for which $2.5\%$ of all the pixels contain more rays and for which $2.5\%$ of all the pixels contain fewer rays, and transform them into magnitudes. We designate these limits by $\Delta_{++}$ and $\Delta_{--}$ respectively.\footnote{We use $\Delta_{++}$ and $\Delta_{--}$ to refer to the cutoffs as well as to the number of rays, the magnification, and the magnitude at those cutoffs, for ease of reference. Which quantity is meant at a particular time should be evident by context.} The bottom portion of Figure \ref{fig:mag_map} shows the histograms for the magnification maps of the top portion, with the values of $\Delta_{++}$ and $\Delta_{--}$ marked. For every point sampled in $(\kappa, s_\star)$ space, we run 10 micro-lensing simulations and calculate the average $\Delta_{++}$ and $\Delta_{--}$ from these 10 simulations. 

\subsection{The Mass-Sheet Degeneracy}

The mass sheet degeneracy \citep{1985ApJ...289L...1F} allows for an arbitrary scaling of lens mass distributions without affecting any of the observable quantities like position and magnification. In the context of micro-lensing, this degeneracy allows for any triplet of $(\kappa,\gamma,s_\star)$ to be converted into an effective doublet\footnote{The choice of doublet is arbitrary. A common alternative is to choose $(\kappa=\kappa_\star, \gamma)$.} through a judicious choice of effective $\kappa_s'$, reducing the parameter space by one dimension while scaling the magnifications and source plane coordinates \citep{1986ApJ...301..503P, 2004ApJ...605...58K, 2014ApJS..211...16V, 2014ApJ...793...96S}. Our simulations, which sample the parameter space of $(\kappa=\gamma,s_\star)$ within the square where $0<\kappa<1$ and $0<s_\star\leq1$, are immediately applicable to the SIEP model. However, they are still relevant for a wider range of parameter values when appropriately scaled. More specifically, given values for $(\kappa, \gamma, s_\star)$, the transformations

\begin{equation}
    1-\kappa' =\frac{1-\kappa}{1-\kappa+\gamma},
\end{equation}
\begin{equation}
    \gamma' = \frac{\gamma}{1-\kappa+\gamma} = \kappa',
\end{equation} and \begin{equation}
    s_\star' = \frac{s_\star \kappa}{\gamma}
\end{equation} allow comparison with our results.

\section{Results} \label{sec:results}

We created contour plots of our averaged $\Delta_{++}$ and $\Delta_{--}$ as functions of $(\kappa,s_\star)$.\footnote{Contour plots were created using Mathematica's ListContourPlot function. They are simple linear interpolations of the data points provided.} The left half of Figure \ref{fig:delta++_delta--} shows $\Delta_{--}$, while the right half shows $\Delta_{++}$. The ``+" symbols on the plots denote the points in the parameter space for which we ran micro-lensing simulations. Data points are symmetrical around $\kappa=0.5$ for the majority of the plots, with the symmetry broken closer to $\kappa=0.5$ to explore regions of interest. The top portions of the plots show the entirety of the parameter space sampled, while the bottom portions of the plots shows a zoom around $\kappa=0.5$ for low values of $s_\star$.

\subsection{$\Delta_{++}$}
\begin{figure*}
    \centering
    \includegraphics[width=\textwidth]{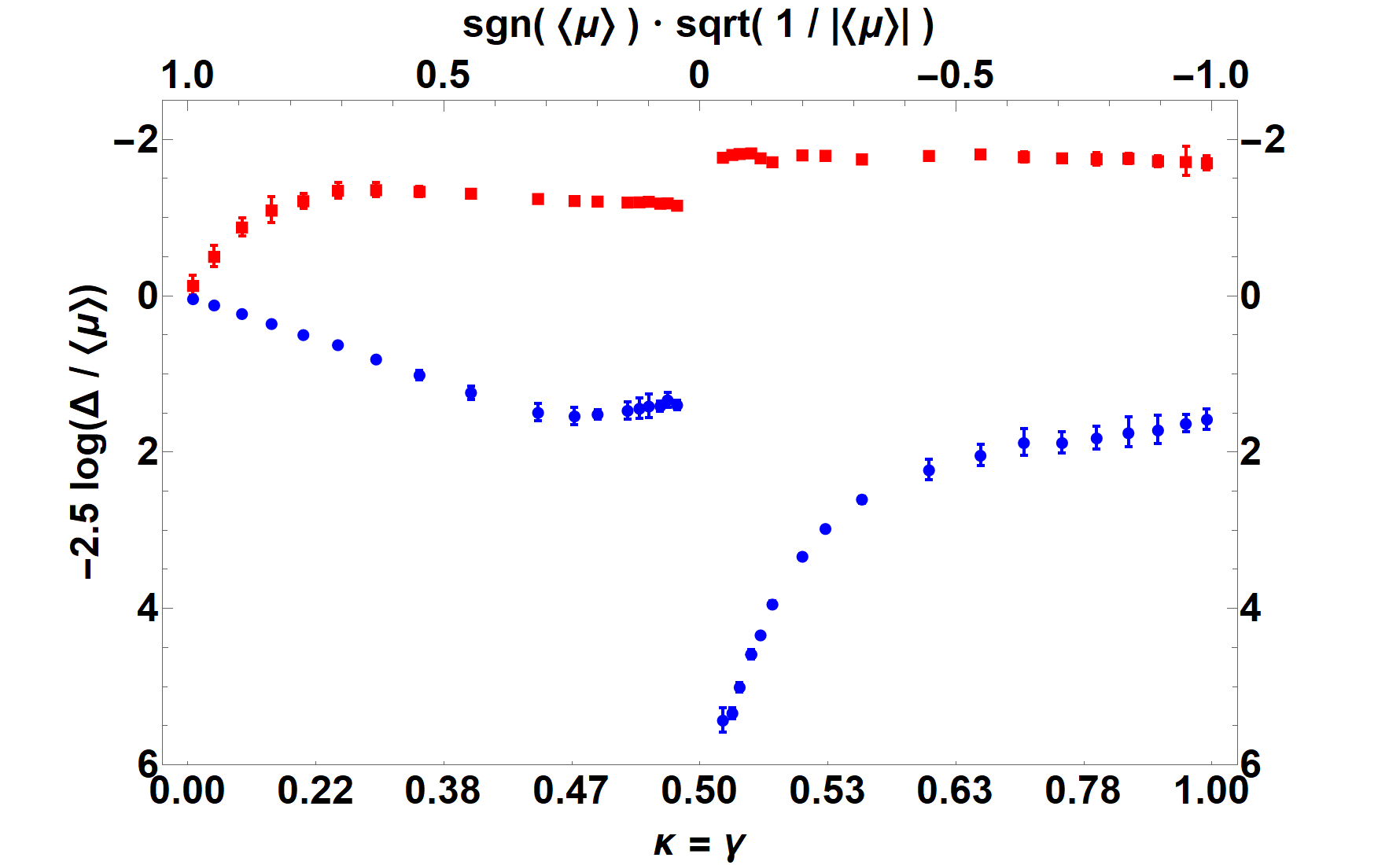}
    \caption{Extremal values from our simulations for (blue, circles) $\Delta_{--}$ and (red, squares) $\Delta_{++}$ as functions of $\kappa=\gamma$. Error bars (most of which are small, compared to the symbols) depict the values of $\sigma_{++}$ and $\sigma_{--}$ for the set of 10 simulations corresponding to each point sampled. The values of $s_\star$ corresponding to the extremal values depicted here can be found in Figure \ref{fig:delta++_delta--}.}
    \label{fig:delta++_delta--_extrema}
\end{figure*}

The plot of $\Delta_{++}$ displays several interesting properties. First, for both macro-minima (the left half, $0<\kappa<0.5$) and macro-saddles (the right half, $0.5<\kappa<1$), there appears to be a ridge along which the value of $\Delta_{++}$ is extremized. For each respective macro-image type, the brightest value of $\Delta_{++}$ appears to be constant along this ridge. For macro-minima we found this value to be $\Delta m\approx -1.3$ mag, while for macro-saddles we found $\Delta m\approx -1.7$ mag. The dashed black line on the plot denotes the locus where $\kappa_\star = 1 / |\mu_{macro}|$, which traces the ridge of brightest $\Delta_{++}$. Figure \ref{fig:delta++_delta--_extrema} displays the values of $\Delta_{++}$ along this ridge.

\subsection{$\Delta_{--}$}
The left half of Figure \ref{fig:delta++_delta--} shows the contour plot for $\Delta_{--}$. Note the difference in the range of magnitudes shown compared to that of $\Delta_{++}$. The contours of this plot are markedly different than that of $\Delta_{++}$, and display different features for macro-minima as opposed to macro-saddles. For the macro-minima, there again appears to be a ridge along which the value of $\Delta_{--}$ is extremized. We find this extremal value to be $\Delta m\approx1.5$ mag. The dashed black line no longer describes where this ridge is located though -- the ridge appears to have some other locus describing it. 

The macro-saddles have no such ridge where $\Delta_{--}$ is extremized; instead, values form a `basin' with slowly sinking contours that grow fainter and fainter. However, for a given vertical cut at a specific $\kappa$ value, the value of $\Delta_{--}$ grows fainter as $s_\star$ increases before reaching an extremum and growing brighter again. This faintest value for a specific $\kappa$ is described once more by the locus where $\kappa_\star = 1/|\mu_{macro}|$. As $\kappa\rightarrow0.5$ and $s_\star\rightarrow0$, the contours of $\Delta_{--}$ slowly sink down further and further, the dimmest possible fluctuations becoming fainter and fainter as the macro-image becomes brighter and the stars farther apart. We were limited by computer power and time, but we found values of the demagnification up to $\Delta m\approx 5.4$ mag in this regime. 

\subsection{Simulation Errors and Uncertainties}

Appendix \ref{app:sim_errors_uncs} contains a discussion of some of the errors associated with micro-lensing simulations, our mitigations of such errors, and the uncertainties in our measured values for $\Delta_{++}$ and $\Delta_{--}$.

\section{Observed ``Worst'' Cases}\label{sec:observed_worst_cases}

``Worst-case" scenarios require the confluence of three factors: the random configuration of micro-lensing stars, the ratio of stellar to total mass surface density, and the ratio of the half-light radius of the source to the Einstein ring radius of the micro-lenses. Likelihoods for the last two of these will vary from one circumstance to another. There are many quadruply lensed sources for which the observed flux ratio anomalies at one or another wavelength are much smaller than the worst case, either by virtue of the source size or the micro-lens surface density. In this section we examine three examples of lensed systems that are consistent with a worst case scenario.

\subsection{Quad Lens RX J1131-1231}

The quad RX J1131-1231 was the thirteenth and last ranked GraL system described in \S1.  We have used Keeton's {\tt lensmodel} program (\citeyear{2001astro.ph..2340K}) to model HST positions for the four images as a singular isothermal sphere with external shear. 

Minimizing the RMS residuals of Gaia magnitudes from the model predictions, we find that image C, the faintest of the cusp images, is 0.34 magnitudes fainter than predicted. This is only 1/4 of the $\Delta_{--}$ excursion shown in Figure \ref{fig:delta++_delta--}. Image D, isolated on the far side of the lens, is 1.20 magnitudes brighter than predicted. This is roughly 3/4 of the $\Delta_{--}$ excursion shown in \ref{fig:delta++_delta--}. 

These fluctuations are not extreme. It raises the question of why the other twelve known lenses exhibited fluctuations that were so much smaller. The widely appreciated answer is that the optical continuum emitting regions of quasars are at least as extended as the Einstein rings of micro-lensing stars. \cite{2007ApJ...661...19P} show that the rms deviations of optical fluxes from those predicted by lens models only are half as large as those observed in X-ray fluxes, which have been found to arise from smaller regions than the optical \citep{2008ApJ...689..755M}. 

\subsection{Quad Lens SDSS J0924+0219}

\cite{2006ApJ...639....1K} call SDSS J0924+0219 ``the most anomalous lensed quasar''.  Using the Keeton et al. model, with the HST V band source flux determined from the B and C images, we find the D image is 2.79 magnitudes fainter than predicted, compared with a $\Delta_{--}$ excursion of 3.1 mag found from Figure 3. Using the X-ray fluxes from \cite{2007ApJ...661...19P} again calibrated by B and C, we find image D is 2.58 magnitudes fainter than predicted. 

One might wonder whether the faintness of image D is due to substructure milli-lensing, but \cite{2020MNRAS.496..138B} have found radio flux ratios that are consistent with the model predictions, ruling out the milli-lensing hypothesis. SDSS J0924+0219 would appear to be a quad in which our worst case estimate is borne out.

\subsection{SN Ia iPTF16geu}

\cite{2020MNRAS.496.3270M} present a model for the observations of the quadruply lensed supernovae iPTF16geu. Though slightly shallower than isothermal, we use their magnifications to compute the isothermal effective $\kappa$.  Their image 1 is 1.00 mag brighter than the model prediction and their image 4 is 0.70 mag fainter than predicted. Their estimated stellar fraction is  $s_\star \sim 0.2$.  Figure \ref{fig:delta++_delta--} give $\Delta_{++} = -1.5$ mag and $\Delta_{--} = 1.3$ mag.  Taken together the fluctuations are roughly what one would expect if the supernova were pointlike at the time of observation.

\section{Conclusions} \label{sec:conclusions}

We have conducted an analysis of the effects of micro-lensing on the brightness of gravitationally lensed images. We used micro-lensing maps to obtain ``worst-case'' uncertainties (which we defined as the boundaries of the 95\% confidence range) in the micro-magnification for different combinations of the convergence $\kappa$ and stellar fraction $s_\star$ of lenses characterized by isothermal potentials where $\kappa=\gamma$. On the faint end ($\mu<\Delta_{--}$), demagnification can be attributed to the absence of extra image pairs -- a macro-minimum has only one micro-minimum and a saddle-point for every star, while a macro-saddle has one micro-saddle corresponding to the global saddle-point and fainter micro-saddles for every star. The fact that minima must be of at least unit magnification while saddle-points can be arbitrarily demagnified leads to some of the differences between the left and right half of the $\Delta_{--}$ plot in Figure \ref{fig:delta++_delta--}. At the bright end ($\mu>\Delta_{++}$), high magnifications are largely due to caustic crossing events or passage near cusps. The bottom half of Figure \ref{fig:mag_map} shows the magnification maps of the top half converted into a 3 color scheme: black for $\mu<\Delta_{--}$, white for $\mu>\Delta_{++}$, and gray for our 95\% interval in between. The caustics largely occur within the region of $\mu>\Delta_{++}$, and would be completely recovered in white in the figure if pixel resolution was increased. 

While our purpose was determining the ``worst-case'' micro-lensing fluctuations, we can comment on the cause of deviations within the 95\% confidence range as well. Deviations within can, for the most part, be attributed to variations in the number of micro-minima. \cite{2003ApJ...583..575G} assert that these variations are greatest when the average number of extra image pairs $\langle n\rangle\sim1$. If the average area of the caustic due to a star scales with $|\mu_{macro}|$ then the covering factor of the caustics, which gives the expected number of extra image pairs, that maximizes the variations can be roughly found as $\kappa_\star\cdot|\mu_{macro}|\sim1$, which recovers the locus $\kappa_\star=1/|\mu_{macro}|$. Alternatively, one can consider the contributions of smooth and grainy matter to the magnification tensor. The stars introduce fluctuations into the magnification tensor on top of the smooth matter component. One might expect these fluctuations to be most impactful when $1-\kappa_s-\gamma=0$, which (for macro-saddles at least) recovers $\kappa_\star=1/|\mu_{macro}|$ as well.

The appendix of \citet{2015ApJ...800...11L} examines the role of $s_\star$ in the RMS fluctuations of magnification maps for a macro-minimum and a macro-saddle. For their macro-minimum, which has $\kappa=0.475$ and $\gamma=0.425$, we would expect the the fluctuations to peak at $\kappa_\star=1/|\mu_{macro}|\approx0.1$. Accounting for the somewhat sparse sampling in their Figure 14, we find this value of $\kappa_\star$ to be consistent with their results. Similarly for their macro-saddle, which has $\kappa=0.475$ and $\gamma=0.425$ resulting in the same macro-magnification, a peak in the RMS near $\kappa_\star\approx0.1$ is clearly seen. Indeed Figure 14 of \citet{2015ApJ...800...11L}, after appropriate conversion using the mass-sheet degeneracy, bears a striking resemblance to a vertical cut of $\Delta_{++}$ or $\Delta_{--}$ in our Figure \ref{fig:delta++_delta--}  at $\kappa=\gamma=0.45$ or $0.55$ (for the minimum or saddle respectively).

Under the worst set of conditions corresponding to micro-lensing of a highly magnified saddle with low stellar surface mass density, our analysis showed that micro-lensing introduces uncertainties of at least three magnitudes, with the uncertainty appearing to increase unbounded as $\kappa\rightarrow 0.5$ and $s_\star\rightarrow 0$. Elsewhere, (de)magnifications appear to be bounded. Fluctuations in the magnification for macro-minima and macro-saddles peak when the stellar fraction $\kappa_\star=1/|\mu_{macro}|$. 

Simple comparisons of observed magnitude differences in lensed systems to the ``worst-case'' fluctuations in Figure \ref{fig:delta++_delta--} can serve as a guide in determining whether more detailed micro-lensing analyses are worthwhile. 

\acknowledgments

Sahil Pontula would like to thank the MIT Undergraduate Research Opportunities Program for support. We would like to thank Jeffrey Blackburne for providing micro-lensing maps and helpful suggestions during the beginning of this work. We would like to thank Georgios Vernardos for providing micro-lensing maps and suggestions at the start of this work as well, and additionally for a thorough readthrough and very helpful comments on the manuscript. We would like to thank Jordi Miralda-Escude and Joachim Wambsganss for their careful reading and comments on the manuscript as well.

Data from this work can be made available via reasonable request to the corresponding author.

\newpage

\appendix

\section{Simulation Errors and Uncertainties}\label{app:sim_errors_uncs}

Any micro-lensing ray tracing simulations are subject to the effects of using a finite number of stars. Rays passing close to a star are deflected out of the receiving square, but there are no stars far away which deflect rays in. \cite{1986ApJ...306....2K} and \cite{1987A&A...171...49S} provide expressions for the number of stars seen in a solid angle containing a desired percentage of the total flux; our code uses enough stars and appropriately sized shooting regions to account for 99\% of the flux.

Any given realizations of a star field will still exhibit slight differences from one another as well. Our decision to average over 10 simulations for each point in $(\kappa,s_\star)$ attempts to mitigate such differences. Figure \ref{fig:delta++_delta--} shows the average $\Delta_{++}$ and $\Delta_{--}$ from each set of 10 simulations. Here, we discuss the spread of each set. 

\begin{turnpage}
    \begin{figure*}
        \centering
        \includegraphics[width=0.48\textheight]{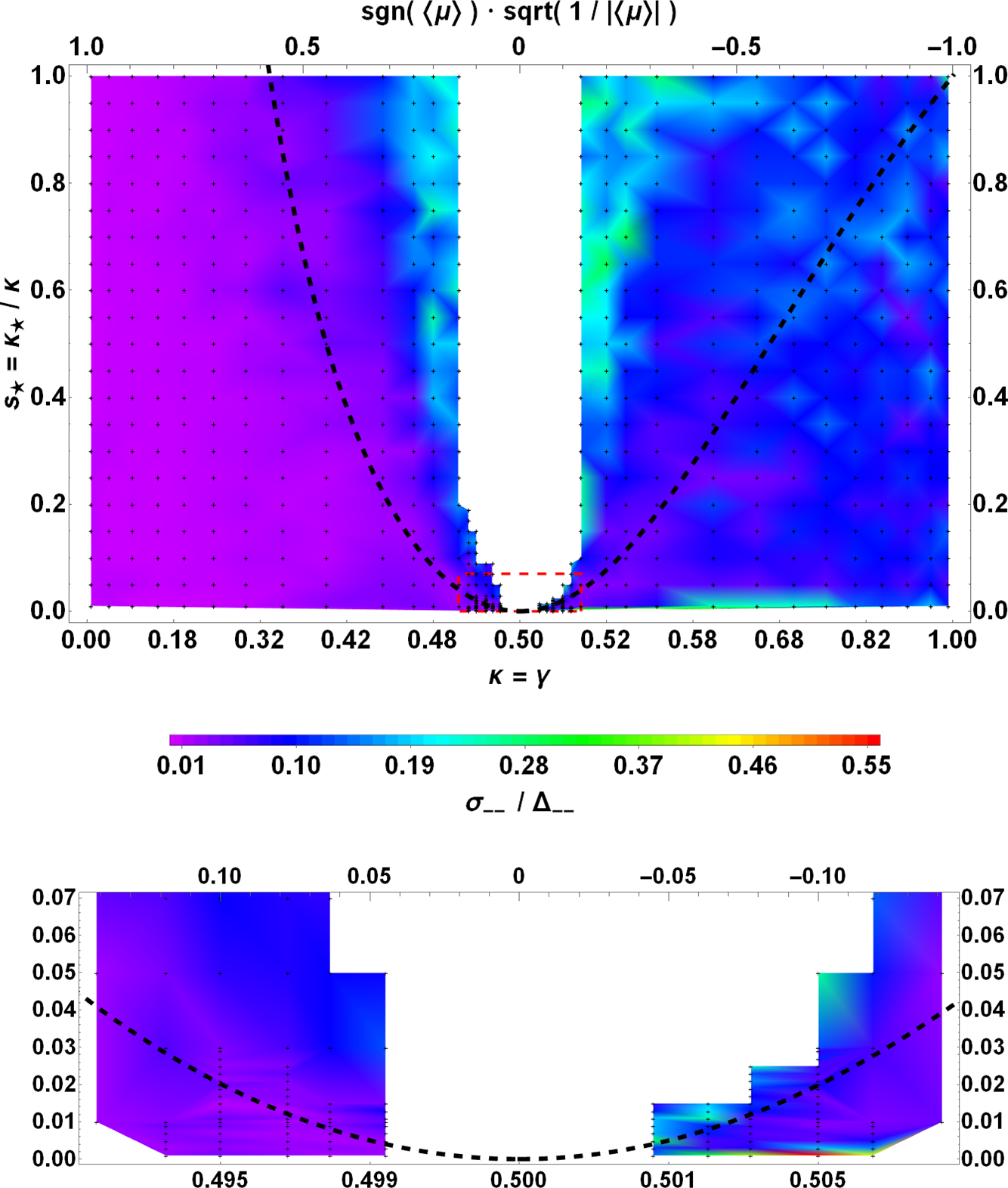}
        \hspace{0.5cm}
        \includegraphics[width=0.48\textheight]{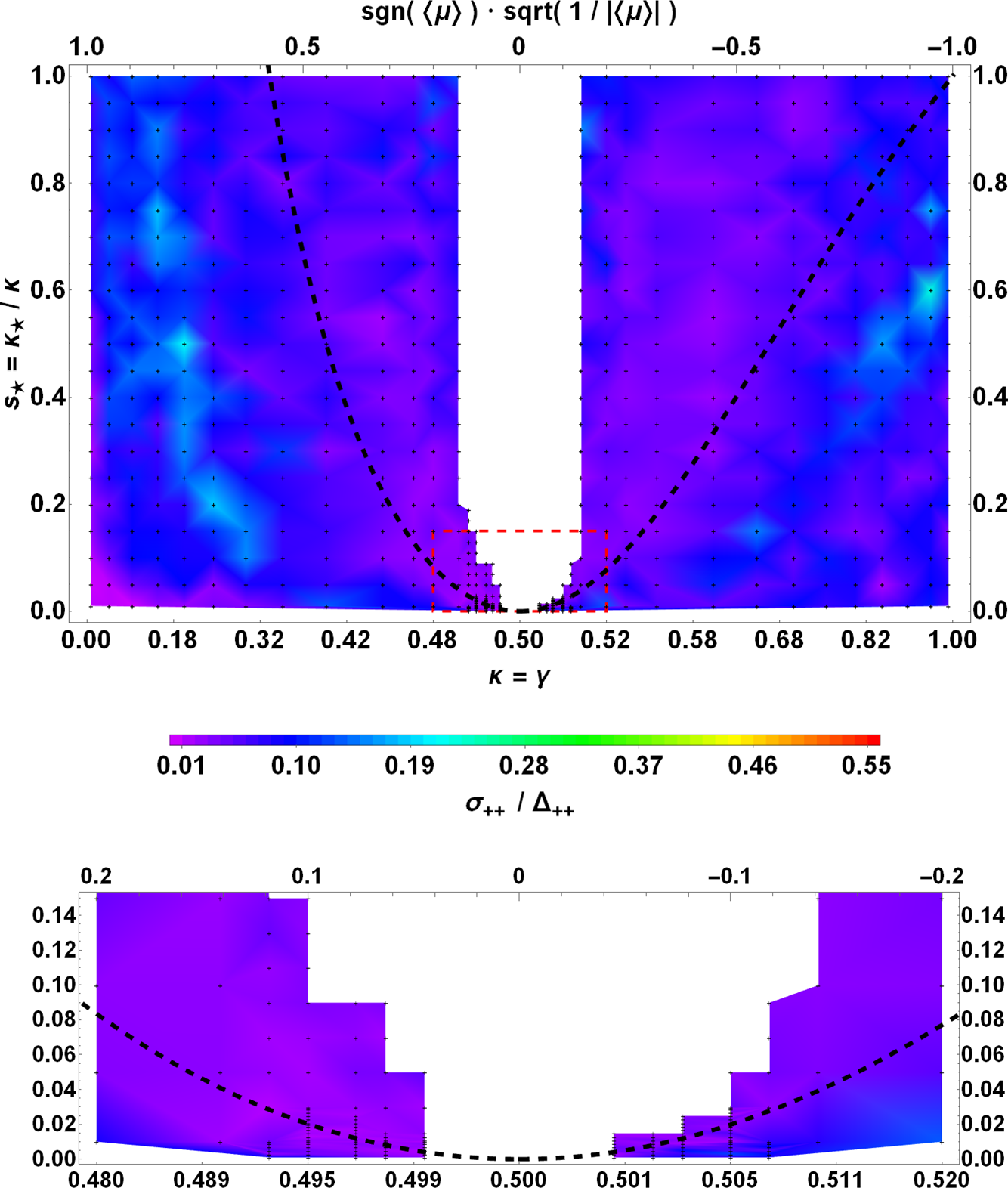}
        \hspace{0.5cm}
        \caption{The uncertainty $\sigma_{--}$ in the number of rays corresponding to $\Delta_{--}$ measured relative to $\Delta_{--}$ (left), and the uncertainty $\sigma_{++}$ in the number of rays corresponding to $\Delta_{++}$ measured relative to $\Delta_{++}$ (right).}
        \label{fig:delta++_delta--_errs}
    \end{figure*}
\end{turnpage}

What we actually measure are the number of rays corresponding to our $\Delta_{++}$ and $\Delta_{--}$ limits in each simulation, so the standard deviation $\sigma$ of each set of 10 simulations is found in units of the number of rays as well. We must then account for the fact that $\Delta_{--}$ is demagnified and will have fewer rays than the magnified $\Delta_{++}$ -- a standard deviation $\sigma_{--}$ of 10 rays for $\Delta_{--}$ is not necessarily better than a standard deviation $\sigma_{++}$ of 100 rays for $\Delta_{++}$. The obvious comparators for $\sigma_{++}$ and $\sigma_{--}$ are the number of rays giving the $\Delta_{++}$ or $\Delta_{--}$ limits themselves. Figure \ref{fig:delta++_delta--_errs} shows density plots of the uncertainties in the number of rays for $\Delta_{++}$ and $\Delta_{--}$ measured relative to themselves. We find that $\sigma_{--}\leq0.1\Delta_{--}$ for the majority of the parameter space covered. This increases slightly to $\sigma_{--}\leq0.3\Delta_{--}$ closer to $\kappa\approx0.5$, particularly for higher values of $s_\star$. There are some additional instances where $\sigma_{--}\approx0.5\Delta_{--}$ for much lower values of $s_\star\approx0.001$, but these can be attributed to the low surface mass density and the fact that our source plane region of $25\theta_E$ is likely somewhat smaller than necessary to adequately sample the caustics. We find that $\sigma_{++}\leq0.23\Delta_{++}$ for the entirety of the parameter space, with the majority having $\sigma_{++}\leq0.1\Delta_{++}$. 

We note here that our choice of $\langle n_{rays}\rangle=1000$ plays a role in the errors. For low macro-magnifications, this choice of $\langle n_{rays}\rangle$ is more than adequate, but for high macro-magnifications a larger number of rays may have been more appropriate (at the expense of more computing time). Some of our magnification maps contained pixels with zero rays, although unsurprisingly this only occurred in the high macro-magnification, low stellar density regime of the macro-saddles where there is the possibility of strong demagnification. Increasing the average number of rays would have removed any 0 count pixels and decreased our uncertainties $\sigma_{--}$ relative to $\Delta_{--}$.

\newpage

%% For this sample we use BibTeX plus aasjournals.bst to generate the
%% the bibliography. The sample63.bib file was populated from ADS. To
%% get the citations to show in the compiled file do the following:
%%
%% pdflatex sample63.tex
%% bibtext sample63
%% pdflatex sample63.tex
%% pdflatex sample63.tex

\bibliography{bibliography}{}
\bibliographystyle{aasjournal}

%% This command is needed to show the entire author+affiliation list when
%% the collaboration and author truncation commands are used.  It has to
%% go at the end of the manuscript.
%\allauthors

%% Include this line if you are using the \added, \replaced, \deleted
%% commands to see a summary list of all changes at the end of the article.
%\listofchanges

\end{document}